\documentclass[12pt]{article}%
\usepackage{amsmath,latexsym}
\usepackage{graphicx}
\usepackage{amsmath}
\usepackage{amsfonts}
\usepackage{amssymb}%
\begin{document}

\title{{Traversable wormholes and the Brouwer
   fixed-point theorem}}
   \author{
Peter K. F. Kuhfittig*\\
\footnote{E-mail: kuhfitti@msoe.edu}
 \small Department of Mathematics, Milwaukee School of
Engineering,\\
\small Milwaukee, Wisconsin 53202-3109, USA}

\date{}
 \maketitle

\begin{abstract}\noindent
The Brouwer fixed-point theorem in topology
states that for any continuous mapping $f$
on a compact convex set into itself admits
a fixed point, i.e., a point $x_0$ such that
$f(x_0)=x_0$.  Under suitable conditions, this
fixed point corresponds to the throat of a
traversable wormhole, i.e., $b(r_0)=r_0$ for
the shape function $b=b(r)$.  The possible
existence of wormholes can therefore be
deduced from purely mathematical
considerations without going beyond the
existing physical requirements.    \\

\noindent
Keywords\\
Traversable Wormholes, Brouwer Fixed-Point Theorem\\
\end{abstract}

\section{Introduction}\label{S:Introduction}

Wormholes are handles or tunnels in spacetime
connecting widely separated regions of our
Universe or entirely different universes.
Morris and Thorne \cite{MT88} proposed the
following line element for the wormhole
spacetime:
\begin{equation}\label{E:line1}
ds^{2}=-e^{2\Phi(r)}dt^{2}+\frac{dr^2}{1-b(r)/r}
+r^{2}(d\theta^{2}+\text{sin}^{2}\theta\,
d\phi^{2}),
\end{equation}
using units in which $c=G=1$.  Here
$\Phi=\Phi(r)$ is called the \emph{redshift
function}, which must be finite everywhere
to prevent the appearance of an event horizon.
The function $b=b(r)$ is called the \emph{shape
function} since it determines the spatial shape
of the wormhole when viewed, for example, in an
embedding diagram \cite{MT88}.  The spherical
surface $r=r_0$ is the \emph{throat} of the
wormhole and is characterized by the following
condition: $b(r_0)=r_0$.  Mathematically
speaking, $r=r_0$ is called a \emph{fixed
point} of the function $b=b(r)$ and will play
a key role in our discussion.  For a traversable
wormhole, an important requirement is the
\emph{flare-out condition} $b'(r_0)< 1$;
also, $b(r)<r$ for $r>r_0$.  The flare-out
condition can only be met by violating the
null energy condition (NEC), which states
that
\begin{equation}
  T_{\alpha\beta}k^{\alpha}k^{\beta}\ge 0
\end{equation}
for all null vectors $k^{\alpha}$, where
$T_{\alpha\beta}$ is the stress-energy
tensor.  Matter that violates the NEC is
called ``exotic" in Ref. \cite{MT88}.  In
particular, for the outgoing null vector
$(1,1,0,0)$, the violation has the form
\begin{equation}
   T_{\alpha\beta}k^{\alpha}k^{\beta}=
   \rho +p_r<0.
\end{equation}
Here $T^t_{\phantom{tt}t}=-\rho$ is the energy
density, $T^r_{\phantom{rr}r}= p_r$ is the
radial pressure, and
$T^\theta_{\phantom{\theta\theta}\theta}=
T^\phi_{\phantom{\phi\phi}\phi}=p_t$ is
the lateral pressure.  For completeness,
let us also list the Einstein field
equations:
\begin{equation}\label{E:Einstein1}
  \rho(r)=\frac{b'}{8\pi r^2},
\end{equation}
\begin{equation}\label{E:Einstein2}
   p_r(r)=\frac{1}{8\pi}\left[-\frac{b}{r^3}+
   2\left(1-\frac{b}{r}\right)\frac{\Phi'}{r}
   \right],
\end{equation}
and
\begin{equation}\label{E:Einstein3}
   p_t(r)=\frac{1}{8\pi}\left(1-\frac{b}{r}\right)
   \left[\Phi''-\frac{b'r-b}{2r(r-b)}\Phi'
   +(\Phi')^2+\frac{\Phi'}{r}-
   \frac{b'r-b}{2r^2(r-b)}\right].
\end{equation}

The purpose of this paper is to make use of
fixed-point theory to show that certain
physical conditions imply the possible
existence of traversable wormholes.  To
that end, we need the following special case
of the Brouwer fixed-point theorem:

\textbf{Theorem} \cite{Brouwer}.  Let $f$
be a continuous function from a closed
interval $[a,b]$ on the real line into
itself.  Then $f$ has a fixed point, i.e.,
there is a point $x_0$ such that $f(x_0)
=x_0$.

A function that maps a set into itself is
called a \emph{self-mapping}.
%END OF SECTION

\section{Some consequences of the Brouwer
   fixed-point \\theorem}

Accrding to Ref. \cite{MTW}, the total
mass-energy $M$ of an isolated star is well
defined as long as one retains spherical
symmetry.  In Schwarzschild coordinates,
\begin{equation}\label{E:shape1}
   \text{Total mass-energy inside radius}
   \,\, r\equiv m(r)=\int^r_04\pi (r')^2
   \rho(r')dr',
\end{equation}
where $\rho(r)$ is the energy density.
Moreover, everywhere outside the star,
\begin{equation}
   m(r)=M\equiv\text{total mass-energy in
   the Newtonian limit}.
\end{equation}
We also have from Eq. (\ref{E:Einstein1})
that
\begin{equation}\label{E:shape2}
   b(r)=\int 8\pi (r')^2\rho(r')dr'.
\end{equation}

The line element (page 608 in Ref.
\cite{MTW}) is given by
 \begin{multline}\label{E:line2}
ds^{2}=-e^{2\Phi(r)}dt^{2}+\frac{dr^2}{1-2m(r)/r}
+r^{2}(d\theta^{2}+\text{sin}^{2}\theta\,
d\phi^{2})\\
=-\left(1-\frac{2M}{r}\right)dt^{2
  }+\frac{dr^2}{1-2M/r}
+r^{2}(d\theta^{2}+\text{sin}^{2}\theta\,
d\phi^{2}), \quad\text{for}\, r>R,
\end{multline}
where $R$ is the radius of the spherical
star.  According to Eqs. (\ref{E:shape1})
and (\ref{E:shape2}), $b(r)=2m(r)$.  With
our wormhole spacetime in mind, a more
convenient form of the line element is
\begin{equation}\label{E:line3}
   ds^{2}=-e^{\Phi(r)}dt^{2}+\frac{dr^2}{1-m(r)/r}
+r^{2}(d\theta^{2}+\text{sin}^{2}\theta\,
d\phi^{2}),\quad \text{inside radius}\, r,
\end{equation}
and
\begin{equation}\label{E:line4}
ds^2=-\left(1-\frac{M}{r}\right)dt^{2}
  +\frac{dr^2}{1-M/r}
+r^{2}(d\theta^{2}+\text{sin}^{2}\theta\,
d\phi^{2}), \quad\text{for}\, r>R,
\end{equation}
where $R$ is the radius of the star.  Now
$m(r)$ corresponds to $b(r)$ in line
element (\ref{E:line1}).

To apply the Brouwer fixed-point theorem,
suppose we let $r=r_1$ be a spherical
core, i.e., $0<r_1<R$.  If $\rho(r)$ is
the mass density of the star for $r>r_1$,
then the mass becomes
\begin{equation*}
   m^*(r)=\int^r_{r_1}4\pi (r')^2\rho(r')dr'.
\end{equation*}
Since $m^*(r_1)=0$, the mapping
\[
   m^*:\, [r_1,\infty)\rightarrow [0,\infty)
\]
is not a self-mapping.  So consider a new
 mapping
\begin{equation}\label{E:m(r)}
   m(r)=\int^r_{r_1}4\pi (r')^2\rho(r')dr'
      +K,
\end{equation}
where $K$ is a constant slightly bigger
than $r_1$.  So $m(r_1)=K$.
Since $r=R$ be the radius of the star, we
now draw the important conclusion that
$m(r)$ in Eq. (\ref{E:m(r)}) maps the
closed interval $[r_1,R]$ into itself,
i.e.,
\begin{equation}\label{E:self}
   [K,R^*]\subset [r_1,R],
\end{equation}
where
\begin{equation}
   R^*=m(R)=\int^R_{r_1}4\pi r^2\rho(r)dr
      +K,
\end{equation}
provided that $R$ is large enough so that
$R^*\le R$.  This conclusion can be
illustrated graphically, as shown in Fig. 1.
\begin{figure}[tbp]
\begin{center}
\includegraphics[width=0.8\textwidth]{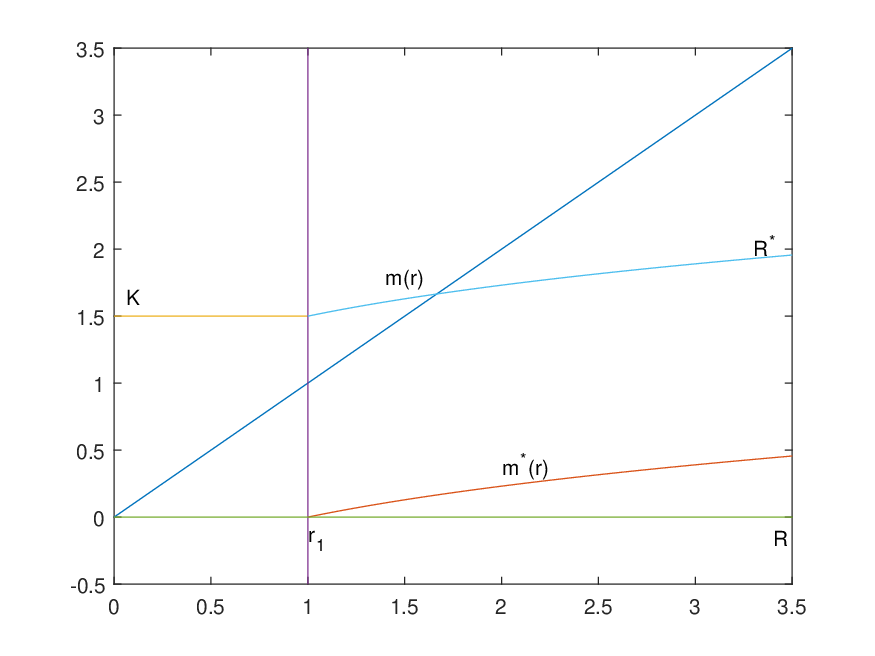}
\end{center}
\caption{$m(r)$ maps the closed interval $[r_1,R]$
   into itself.}
\end{figure}

We also have
\begin{equation}\label{E:mprime}
   m'(r)=4\pi r^2\rho(r)\ll 1,
\end{equation}
since $\rho(r)$ is very small in our
geometrized units.  So while $m(r)$ is
an increasing function, $m'(r)$ remains
less than unity.  (This also explains
why $R^*\le R$ if $R$ is large enough.)

Since $b(r)\equiv m(r)$, we now obtain
from the Brouwer fixed-point theorem that
\begin{equation}
   b(r_0)=m(r_0)=\int^{r_0}_{r_1}
   4\pi r^2\rho(r)dr+K=r_0.
\end{equation}
By Eq. (\ref{E:mprime}), $b'(r_0)
=m'(r_0)<1$, so that the flare-out
condition is satisfied.

In the resulting wormhole spacetime, the
region inside the throat $r=r_0$ is not part
of the wormhole, but this region still
contributes to the gravitational field.
This can be compared to a thin-shell
wormhole resulting from a Schwarzschild
black hole \cite{PV95}: while not part
of the manifold, the black hole generates
the underlying gravitational field.

As actual (quantified) example of the
type of wormhole discussed is given in
Ref. \cite{pK13}.  It is shown that for
a typical neutron star, the possible
formation of a wormhole requires a core
of quark matter that is approximately
1 m in radius.  (Quark matter is
believed to exist at the center of
neutron stars \cite{PSS}.)
%END OF SECTION

\section{A dark-matter background}
In the discussion of dark matter, several
models have been proposed for the energy
density.  The best-known of these is the
Navarro-Frenk-White model \cite{NFW}
\begin{equation}
   \rho(r)=\frac{\rho_s}{\frac{r}{r_s}
   \left(1+\frac{r}{r_s}\right)^2},
\end{equation}
where $r_s$ is the characteristic scale
radius and $\rho_s$ is the corresponding
density.  The Universal Rotation Curve
\cite{CFVS} is given by
\begin{equation}
   \rho(r)=\frac{\rho_cr_c^3}
   {(r+r_c)(r^2+r_c^2)},
\end{equation}
where $r_c$ is the core radius of the galaxy
and $\rho_c$ is the central halo density.
Another example is the King model whose
energy density is given by \cite{iK72}
\begin{equation}
   \rho(r)=\kappa\left(\frac{r^2}{r_0}
     +\lambda\right)^{\eta},
\end{equation}
where $\eta$, $\kappa$, $r_0$, and
$\lambda$ are constants.

All of these models have a low energy
density.  So if $r_1$ is the radius of a
star, then the star itself becomes the
core.  Furthermore, since there is no
outer boundary, we can choose $R$ large
enough so that $m(r)$ in Eq. (\ref{E:m(r)})
is a self-mapping.  The existence of a
fixed point now implies the possible
existence of a wormhole in the
dark-matter region.  This conclusion
is independent of the dark-matter
model chosen.

\emph{Remark:}  The form of $b(r)$ can
be obtained from Eq. (\ref{E:shape2})
in conjunction with one of the above
choices of $\rho(r)$.  The determination
of the integration constant requites an
extra condition, however: a physically
viable assumption is the usual
$b(r_0)=r_0$.  (See, for example, Ref.
\cite{Salucci}.)
%END OF SECTION

\section{Conclusion}

If $r=r_1$ denotes the radius of the core
of the star, then $m(r)$, the effective
mass of the star for $r>r_1$, is given by Eq.
(\ref{E:m(r)}).  The function $m(r)$ satisfies
the hypothesis of the Brouwer fixed-point
theorem.  The fixed point can be viewed as
the radius of the throat of a traversable
wormhole since $b(r_0)= m(r_0)=r_0$ and
$b'(r_0)=m'(r_0)<1$, thereby satisfying the
flare-out condition.  This result can also be
applied to a dark-matter setting by treating
a star of radius $r=r_1$ as the core.  So the
possible existence of traversable wormholes
follows directly from purely mathematical
considerations without going beyond the
physical requirements already in place.

\end{document}